\documentclass[global,twocolumn]{svjour}
\usepackage{latexsym}
\usepackage{graphics}
\usepackage{graphicx}
\usepackage{psfrag}
\usepackage[dvips,colorlinks]{hyperref}

\begin{document}
\title{Loading of large ion Coulomb crystals into a linear Paul trap incorporating an optical cavity for cavity QED studies}
\author{P. Herskind\thanks{E-mail:herskind@phys.au.dk, Fax:+45 8612 0740}, A. Dantan, M. B.
Langkilde-Lauesen, A. Mortensen, J. L. S${\o}$rensen and M.
Drewsen} \institute{QUANTOP, Danish National Research Foundation
Center for Quantum Optics, Department of Physics and Astronomy,
University of Aarhus, DK-8000 \AA rhus C., Denmark}
\date{\today}

\maketitle
%
%
\begin{abstract}
We report on the loading of large ion Coulomb crystals into a
linear Paul trap incorporating a high-finesse optical cavity
($\mathcal{F}\sim3200$). We show that, even though the 3-mm
diameter dielectric cavity mirrors are placed between the trap
electrodes and separated by only 12 mm, it is possible to produce
\textit{in situ} ion Coulomb crystals containing more than 10$^5$
calcium ions of various isotopes and with lengths of up to several
millimeters along the cavity axis. We show that the number of ions
inside the cavity mode is in principle high enough to achieve
strong collective coupling between the ion Coulomb crystal and the
cavity field. The results thus represent an important step towards
ion trap based Cavity Quantum ElectroDynamics (CQED)
experiments using cold ion Coulomb crystals. 
\end{abstract}


\section{Introduction} \label{sec:introduction}
In recent years there has been much focus on light-matter
interactions at the level of single photons, primarily motivated
by applications within quantum information science, where an
efficient light-matter interface is indispensable for many
applications\cite{Cirac_97,Duan_01}. Of particular interest has
been the so-called strong coupling regime of Cavity Quantum
ElectroDynamics (CQED)\cite{Berman}, in which the rate of coherent
evolution within the system exceeds that of any dissipative
processes, making this regime ideal for the quantum engineering of
light-matter interfaces. A strong collective coupling between a
medium of $N$ atoms and the light field can be achieved if the
single atom-photon coupling strength $g_0$ is large enough to
satisfy the inequality $g_0\sqrt{N}>\gamma,\kappa$, where $\gamma$
and $\kappa$ are the decay rates of the atomic dipole and the
cavity field, respectively. For optical frequencies, experiments
with neutral atoms have been extremely successful and have reached
the strong coupling regime for single atoms\cite{Kimble_98}. Such
experiments benefit from the use of cavities of extremely small
mode volumes to increase the single atom-photon coupling strength,
given by:
$g_0=\frac{D}{\hbar}\sqrt{\frac{\hbar\omega}{2\epsilon_0V}}$,
where $D$ is the transition dipole moment, $\omega$ is the
frequency of the transition and $V$ is the mode volume of the
cavity. However, since the cavity decay rate is inversely
proportional to the cavity length, such experiments require very
high finesse cavities, which are technically challenging to
operate. One further complication of such experiments is the
difficulty in efficiently trapping and confining the neutral atoms
inside the cavity. While short-time solutions to this problem have
recently been found\cite{Boozer_06,Puppe_07}, ions, by contrast,
can be stored for hours in ion traps and offer relatively easy
access to a regime in which the spatial extent of their wave
packet is well-localized on the scale of the optical wavelength of
their atomic transition (Lamb-Dicke regime). Though experiments
with single ions in cavities have confirmed
this\cite{Keller_2003,Mundt_2003}, the strong coupling regime has
not yet been reached, mainly due to the difficulty in lowering the
mode volume, i.e. minimizing the mirror separation, without
severely modifying the trapping potential, which makes trapping
extremely challenging. The requirement of a small mode volume can
be relaxed for ensembles of atoms or ions due to the $\sqrt{N}$
factor entering in the collective coupling strength of the
ensemble. This means that a technically less demanding longer
cavity with a lower finesse can be used in such
experiments.\\\indent In addition to the issue of tight
confinement and long storage time, ion Coulomb crystals have a
number of advantages over cold atomic samples. As the ions are
confined in a crystal lattice, the decoherence rate due to
collisions is very low. Furthermore, the lattice structure in
conjunction with the standing wave field of the optical resonator
might be interesting for the engineering of the atom-photon
interaction. Whereas spatial structuring would require the use of
optical lattices for neutral atoms it is inherent in an ion
Coulomb crystal and as described in a recent
paper\cite{Mortensen_07}, the structure of ion Coulomb crystals
can be made very regular and periodic. Finally, the low optical
density of the ion Coulomb crystal ($\sim10^8\mathrm{cm}^{-3}$)
makes optical pumping and state preparation unproblematic.

In this paper we focus on the production and characterization of
large cold ensembles of ions, specifically ion Coulomb crystals,
of up to hundreds of thousands of ions in a linear Paul trap
incorporating a high finesse optical cavity designed for CQED
experiments. We show that it is possible to insert cavity mirrors
in between the trap electrodes without significantly perturbing
the trapping fields, that the ion Coulomb crystals can be produced
in a clean and efficient way that preserves the finesse of the
cavity, and that the number of ions in the cavity mode can be high
enough to potentially satisfy the strong coupling criterion,
$\mathrm{g_0}\sqrt{N}>\gamma,\kappa$. These results thus represent
an important step towards ensemble-based CQED with ion Coulomb
crystals and open up for the use of such media as a tool for
quantum information
processing\cite{Mortensen_thesis,Herskind_proceedings,Coudreau_07}.

The paper is organized as follows: In section
\ref{sec:experimental}, we describe our linear Paul trap along
with the integrated cavity. In section \ref{sec:loading}, we
describe the loading scheme used for the production of the Coulomb
crystals of calcium ions and present results on loading of the
trap with various naturally abundant calcium isotopes. In section
\ref{sec:optimization}, we address the issue of optimization of
the loading in terms of maximizing the total number of ions within
the cavity mode volume. In section \ref{sec:conclusion}, we
conclude.

\section{Experimental setup} \label{sec:experimental}
\begin{figure}\sidecaption
\resizebox{1\hsize}{!}{\includegraphics*{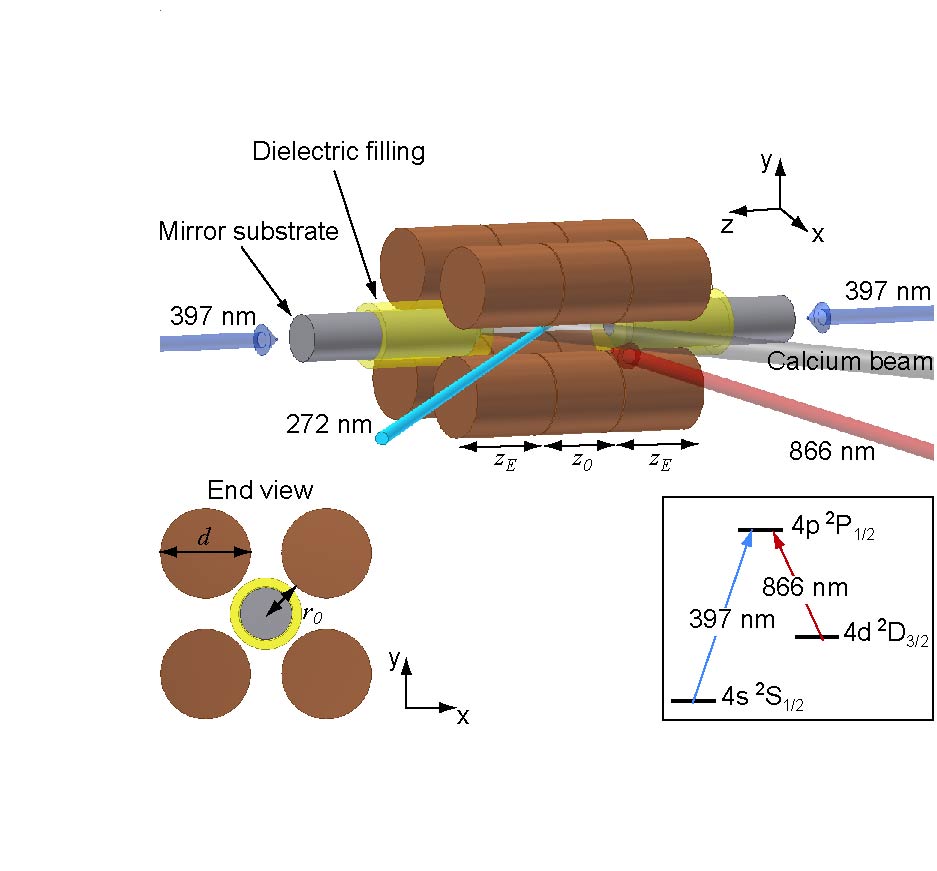}}
\caption{Schematic of the linear Paul trap with laser and atomic
beams indicated. The dielectric fillings have been included to
minimize the rf-fields along the trap axis (z-axis). The insert
shows the level scheme used in Doppler laser cooling of Ca$^+$.}
\label{setup}
\end{figure}

Fig.~\ref{setup} shows a sketch of the setup used in the
experiments. The trap is a linear Paul trap\cite{Prestage_89}
which consists of four sectioned cylindrical electrode rods placed
in a quadrupole configuration. The length of the center-electrode
is $z_0=5.0$ mm, and the length of the end-electrodes is $z_E=5.9$
mm. The electrode diameter is $d=5.2$ mm and the distance from the
trap center to the electrodes is $r_0=2.35$ mm. Confinement in the
radial plane ($xy$-plane in Fig.~\ref{setup}) is obtained by
applying time varying voltages
$\frac{1}{2}U_{rf}\cos(\Omega_{rf}t)$ and
$\frac{1}{2}U_{rf}\cos(\Omega_{rf}t+\pi)$ to the two sets of
diagonally opposite electrode rods. The sectioning of each of the
electrode rods allows for application of a static voltage,
$U_{end}$, to the end-electrodes, which provides confinement along
the $z$-axis. The potential along the $z$-axis is well described
by
\begin{equation} \label{eq:pot_z}
\Phi_z(z)=\frac{1}{2}M\omega_z^2z^2,
\hspace{0.5cm}\omega_z^2=\frac{2\eta QU_{end}}{Mz_0^2}
\end{equation}
where $M$ and $Q$ are the mass and charge of the ion and
$\eta=0.342$ is a constant related to the trap geometry. For
$2QU_{rf}/Mr_0^2\Omega_{rf}^2\ll1$ one can introduce an effective
or pseudo-potential in the radial plane described
by\cite{Drewsen_00}
\begin{equation} \label{eq:pot_r}
\Phi_r(r)=\frac{1}{2}M\omega_r^2r^2, \mbox{ }\mbox{  }\mbox{  }
\omega_r^2=\frac{Q^2U_{rf}^2}{2M^2r_0^4\Omega_{rf}}-\frac{\eta
QU_{end}}{Mz_0^2}.
\end{equation}
In such a three-dimensional potential, one finds that the density
of ions in an ion Coulomb crystal is controlled only by the
rf-voltage and given by\cite{Hornekaer_01}
\begin{equation} \label{eq:density}
\rho=\frac{\epsilon_0U_{rf}^2}{Mr_0^4\Omega_{rf}^2}.
\end{equation}
The trap is operated at a frequency $\Omega_{rf}=2\pi\times4.0$
MHz and the end-voltage, $U_{end}$, is typically 1 to 10 V. The rf
voltage, $U_{rf}$, is typically between 100 and 400 V, which leads
to ion densities between $6.8\times 10^7$ and $1.1\times 10^9$
cm$^{-3}$.

The unique feature of this trap is the incorporation of a high
finesse optical cavity, with mirrors situated in between the
electrodes. The cavity is designed for operating on the
4d~$^2$D$_{3/2}$$\rightarrow$4p~$^2$P$_{1/2}$ transition of
$^{40}$Ca$^+$ at 866 nm (see insert Fig.~\ref{setup}). The
mirrors, both with a radius of curvature of 10 mm, are placed in a
near-confocal geometry supporting a standing wave mode of light at
866 nm with a waist of $\sim 37 \mu$m at the trap center. The
transmission of the two mirrors is 1500 ppm and 5 ppm at 866 nm,
respectively. The cavity finesse at 866 nm is determined by
measuring the free spectral range (FSR) and the width of the
resonance peak. The FSR is measured to be
$\Delta\nu_{\mathrm{FSR}}=(12.7\pm 0.6)$ GHz, corresponding to a
cavity length of 11.8 mm. This value was obtained by tuning the
laser frequency over one FSR of the cavity while recording the
number of (known) FSR of a longer reference cavity. The width of
the resonance peak is found to be $\delta \nu=(4.0 \pm 0.2)$ MHz
by comparing the width with sidebands at 5 MHz. The resulting
finesse is then
$\mathcal{F}=\Delta\nu_{\mathrm{FSR}}/\delta\nu=3200\pm 300$. This
is also consistent with the incoupling mirror transmission of 1500
ppm and intra-cavity losses of 350 ppm, which we derived from the
cavity reflection signal. The intra-cavity loss is due to mirror
contamination introduced mainly during degassing of the ion pump
connected to the vacuum chamber and will be avoided in the future
by inserting a valve between the ion pump and the trap chamber.
This should allow for intra-cavity loss of only a few tens of ppm.
Based on the measurements quoted above, we deduce a cavity decay
rate of $\kappa=2\pi\times(2.0\pm0.1)$ MHz. From the spontaneous
decay rate, $\Gamma$, of the excited 4p~$^2$P$_{1/2}$ state of
$^{40}$Ca$^+$, the decay rate for the atomic dipole is
$\gamma=\Gamma/2=2\pi\times11$ MHz. From the dimensions of the
cavity, the ion-photon coupling strength of a single $^{40}$Ca$^+$
ion at the cavity waist and field anti-node is expected to be
$g_0=2\pi\times0.53$ MHz. In principle it thus requires
$\simeq500$ ions within the cavity mode volume to enter a regime
governed by a strong collective coupling (where
$g_0\sqrt{N}>\gamma,\kappa$ is satisfied).

\section{Loading of the trap} \label{sec:loading}
In order to efficiently load large crystals into the trap we
employ the technique of resonance-enhanced
photo\-ionization\cite{Kjaergaard_00}. As compared to conventional
electron bombardment, this method has several advantages which are
all essential to the experiment. First of all, the problem of
charging surrounding isolating materials, e.g. mirror substrates,
is greatly reduced as only a single electron is produced per
ion\cite{Kjaergaard_00,Brownnutt_07}. Secondly, the ionization
efficiency can be made much higher, leading to loading rates which
are significantly higher than what can be achieved with electron
bombardment\cite{Hendricks_07}. This makes loading at a lower
atomic flux practical\cite{Gulde} and thus reduces the risk of
contaminating both the trap electrodes and closely spaced delicate
objects, such as the mirrors integrated into the ion trap. While
deposition of material on the electrodes is suspected to give rise
to heating of the ions \cite{Deslauriers_06_1,Devoe_02},
contamination of the mirrors will be devastating to the quality of
the resonator. Finally, the technique can be extremely isotope
selective\cite{Mortensen_04} allowing for loading of different
isotopes in well-controlled ratios.

\begin{figure}\sidecaption
\resizebox{1\hsize}{!}{\includegraphics*{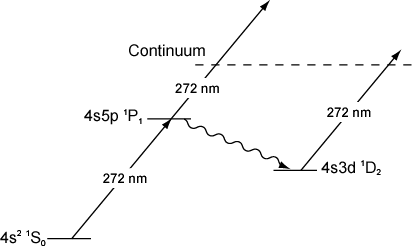}}
\caption{Resonant two-photon ionization scheme used in this work.}
\label{fig1}
\end{figure}

The level scheme used in this work for photoionization is
presented in Fig.~\ref{fig1}. A single UV-light source at 272 nm
ionizes atomic calcium in a two-photon process through resonant
excitation of the 4s5p~$^1$P$_1$ state followed by absorption of a
second 272 nm photon from either the 4s5p~$^1$P$_1$ state or the
4s3d~$^1$D$_2$ state (populated through spontaneous emission) into
the continuum (see Fig.~\ref{fig1})
\cite{Kjaergaard_00,Mortensen_04}. The light at 272 nm originates
from a laser system based on a commercial Ytterbium-doped DFB
fiber laser that has been frequency doubled twice to produce light
at the fourth harmonic. The DFB fiber laser system was described
in detail in a recent paper\cite{Herskind_07}. In brief, light
from a CW DFB fiber laser operating at 1088 nm is first frequency
doubled in a bow-tie cavity containing a LiNbO$_3$ crystal to
produce light at 544 nm. This light is subsequently frequency
doubled in a bow-tie cavity containing a $\beta$-BaBa$_2$O$_4$
crystal to finally produce light at 272 nm. Frequency tuning can
be achieved by controlling the length of the fiber laser cavity
either with a piezoelectric transducer (PZT), or by changing the
temperature.

For loading of the trap, a thermal beam of atomic calcium with a
cross section of $\sim1.0\times1.5$ mm$^2$ is intersected by the
272 nm laser beam in the center of the trap at right angles in
order to minimize Doppler broadening. The atomic beam is produced
from an effusive oven followed by a series of skimmers, inserted
to avoid deposition of calcium on the trap electrodes and cavity
mirrors. The UV-beam has a power of about 20 mW and a waist of
$\sim 160$ $\mu$m at the trap center. The ions produced are then
Doppler laser cooled on the
4s~$^2$S$_{1/2}$$\rightarrow$4p~$^2$P$_{1/2}$ transition along the
trap axis (z axis in Fig.~\ref{setup}) by two counter propagating
beams at 397 nm (beam diameter $\sim$ 1 mm), while in the radial
plane (xy-plane in Fig.~\ref{setup}) the ions are sympathetically
cooled through the Coulomb interaction. An 866 nm beam, resonant
with the 4d~$^2$D$_{3/2}$$\rightarrow$4p~$^2$P$_{1/2}$ transition,
is applied to prevent the ions from being shelved into the
metastable D$_{3/2}$ state (see insert of Fig.~\ref{setup}).
Typically, 8-10 mW of 397 nm light detuned by $\sim 20-40$ MHz
below resonance is used during loading. Both power and detuning
are then decreased upon completion of the loading in order to
optimize the cooling. Detection of the ions is performed by
imaging spontaneously emitted light at 397 nm onto an image
intensified CCD camera located above the trap in Fig.~\ref{setup}.
The number of ions loaded can then be deduced from the recorded
images by the method described in Ref.~\cite{Madsen_00}.

\begin{figure}\sidecaption
\resizebox{1\hsize}{!}{\includegraphics*{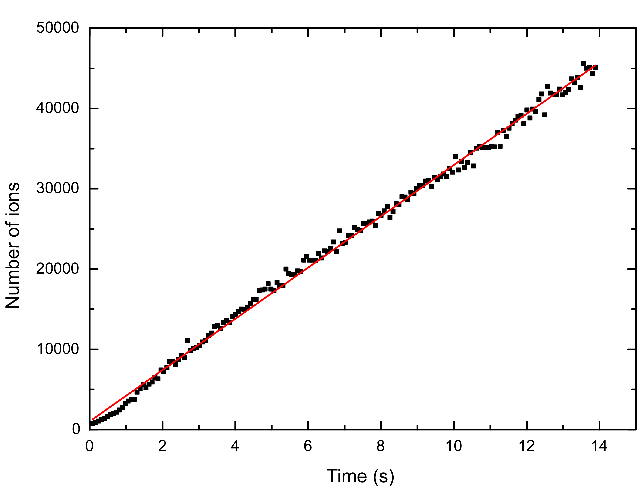}}
\caption{Number of ions loaded versus time when the ionization
laser frequency is tuned close to the
4s$^2$~$^1$S$_0$$\rightarrow$4s5p~$^1$P$_1$ transition of
$^{40}$Ca. The UV-power used for ionization was 20 mW and the oven
temperature 400$^{\circ}$C. The loading rate deduced from the
linear fit is $\sim 3200$ ions/s.} \label{fig3}
\end{figure}
\begin{figure}\sidecaption
\resizebox{1\hsize}{!}{\includegraphics*{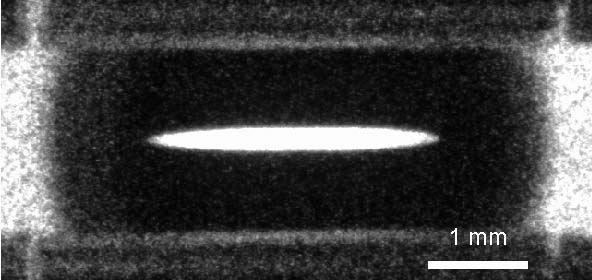}}
\caption{Image of a 3 mm long $^{40}$Ca$^+$ Coulomb crystal with
88000 ions. The rf and end-voltages are 300 V and 1.7 V,
respectively. The visible features outside the crystal region are
from the trap and due to light scattered from the 397 nm beams as
they pass through the mirrors. Due to the low magnification needed
to image the whole crystal, no detailed crystal structure is
visible.} \label{largesinglecrystal}
\end{figure}

Fig.~\ref{fig3} shows the number of loaded ions versus time when
the frequency of the 272 nm light source is tuned close to the
resonance of the 4s$^2$~$^1$S$_0$$\rightarrow$4s5p~$^1$P$_1$
transition of $^{40}$Ca. The end- and rf-voltages were
$U_{end}$=3.9 V and $U_{rf}$=130 V, respectively
($\omega_r=2\pi\times225$ kHz and $\omega_z=2\pi\times160$ kHz),
and the oven temperature during the loading sequence was
$400^{\circ}$C. As can be seen from the figure, we are able to
load in the excess of 3000 ions/s at this relatively low oven
temperature. This means that Coulomb crystals with more than
$10^5$ ions can be produced within a minute.
Fig.~\ref{largesinglecrystal} shows an example of such a crystal,
where the total number of ions is 88000 and the density and length
of the crystal are $6.1\times10^8\mathrm{cm}^{-3}$ and 3 mm,
respectively.

\begin{figure}\sidecaption
\resizebox{1\hsize}{!}{\includegraphics*{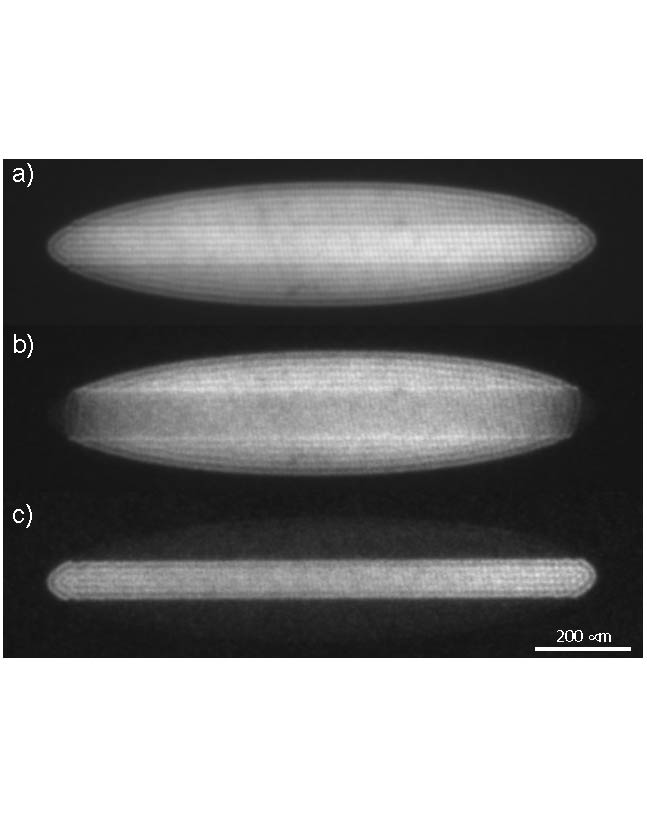}}
\caption{a) Image of a two-component crystal consisting of $\sim
2000$ $^{40}$Ca$^+$ (inner component) and $\sim 13000$
$^{44}$Ca$^+$ (outer component) that are being laser cooled
simultaneously (16 s exposure time). b) $^{44}$Ca$^+$ laser
cooled, $^{40}$Ca$^+$ sympathetically cooled (1 s exposure time).
c) $^{40}$Ca$^+$ laser cooled, $^{44}$Ca$^+$ sympathetically
cooled (1 s exposure time). For technical reasons the cooling
light resonant with the sympathetically cooled component was not
turned off completely and in both b) and c) this component is
still weakly visible.} \label{fig:Ca44_inner_outer_both}
\end{figure}

\begin{figure}\sidecaption
\resizebox{1\hsize}{!}{\includegraphics*{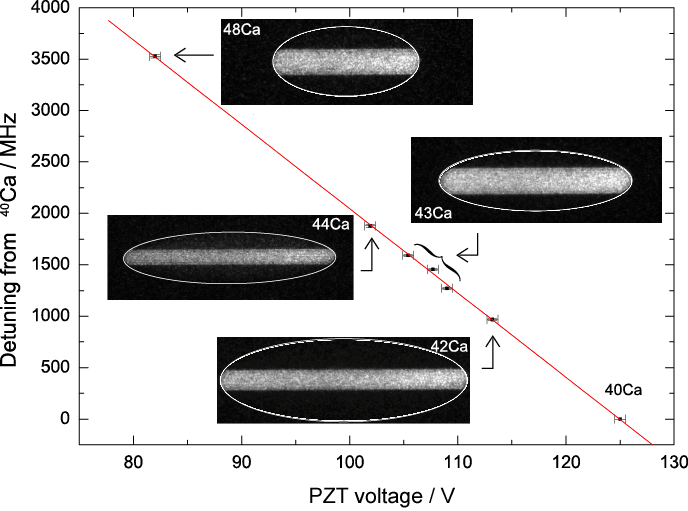}}
\caption{Frequency detuning with respect to the $^{40}$Ca
resonance vs. voltage applied to the fiber laser PZT. The three
points around $^{43}$Ca correspond to the different hyperfine
splitting of the 4s5p$^1$P$_1$ level and the frequency shifts of
the various isotopes with respect to the $^{40}$Ca resonance are
taken from Ref.~\cite{Mortensen_04}. The fit indicates a frequency
shift of $82\pm2$ MHz/V at 272 nm and the error bars reflect the
uncertainty in determining the peak of the ionization resonance.
Inserts show typical two-component crystals of the various
isotopes, for the same rf-voltage of 220 V, but unequal
end-voltages. The dimensions of the black frames are 1.4 mm
$\times $ 0.5 mm and the ellipsoids indicate the boundaries of the
outer component of the two-component crystals.}
\label{fig:pztCalibrationGraph2}
\end{figure}

A potentially attractive feature of ion Coulomb crystal based CQED
is the possibility to work with crystals of various isotope
contents. Two-component crystals are especially interesting for
CQED studies since they allow for laser cooling one component
(outer, radially separated component) while having the other
component (inner cylindrical component) interacting with the
cavity field only. By tuning our 272 nm light source to resonance
with the respective transitions of specific isotopes of calcium,
we have loaded such two-component crystals consisting of
$^{40}$Ca$^+$ and isotopes of higher mass numbers. When producing
$^{\mathrm{M}}$Ca$^+$, where $\mathrm{M}>40$, $^{40}$Ca$^+$ ions
are also created through an electron charge transfer process
between atoms in the atomic beam, dominated by $^{40}$Ca, and the
ions in the trap\cite{Mortensen_04}. This process has the form,
$^{40}$Ca+$^{\mathrm{M}}$Ca$^+$ $\rightarrow$
$^{40}$Ca$^+$+$^{\mathrm{M}}$Ca+$\Delta$E$^{\mathrm{M}}$ and is
nearly resonant in the sense that $\Delta E^{\mathrm{M}}$ lies
much below the energy associated with the thermal collisions
leading to the exchange process. The relative content of
$^{40}$Ca$^+$ and less naturally abundant isotopes in the crystal
can be controlled to a high degree by turning on the atomic beam
after the ionization laser has been turned off. Due to the mass
dependence of the radial trapping potential (c.f.
eq.~\ref{eq:pot_r}) the different isotopes separate radially when
cooled into an ion Coulomb crystal\cite{Hornekaer_01}. This is
clearly seen in Fig.~\ref{fig:Ca44_inner_outer_both}a) where both
$^{40}$Ca$^+$ and $^{44}$Ca$^+$ ions are laser cooled
simultaneously. In Fig.~\ref{fig:Ca44_inner_outer_both}b) and c)
the cooling lasers have been applied only to one component, while
the other is being sympathetically cooled.
Two-component crystals of $^{40}$Ca$^+$ ions and all other
naturally abundant isotopes of calcium (except $^{46}$Ca$^+$ which
has a natural abundance of only $\sim10^{-5}$\cite{Emsley_95})
have been produced through the charge transfer process described
above, as seen in Fig.~\ref{fig:pztCalibrationGraph2}. The laser
system is capable of covering the entire spectrum of naturally
abundant calcium and allows for easy and quick changing from one
isotope to another, which makes CQED studies with such
two-components crystals practical. The relevant transition for the
CQED experiments on this system is the
4d$^2$D$_{3/2}$$\rightarrow$4p$^2$P$_{1/2}$ transition of Ca$^+$.
The isotope shift of this transition, relative to $^{40}$Ca$^+$,
is 4.5~GHz for $^{44}$Ca$^+$, which means that $^{40}$Ca$^+$
should not be affected by the cooling lasers when this component
is being sympathetically cooled by the $^{44}$Ca$^+$ ions.
However, with our loading scheme we have the possibility to work
with $^{48}$Ca$^+$ instead of $^{44}$Ca$^+$ which has an even
larger isotope shift of 8.3~GHz.\cite{Nortershauser_98}

As mentioned previously, compared to electron bombardment, the
method of resonant photoionization minimizes charging effects as
well as trap contamination and formation of patch potentials. From
the images shown in Fig.~\ref{fig:Ca44_inner_outer_both} and
\ref{fig:pztCalibrationGraph2}), there appears to be no visible
perturbation on the shape caused by such effects. In
Fig.~\ref{fig:Ca44_inner_outer_both} the upper and lower
boundaries of the inner component have been measured to be
parallel at least to within 2 mrad, indicating that the
introduction of cavity mirrors inside a linear Paul trap does not
significantly perturb the trapping of such crystals. Furthermore,
as the high-finesse cavity constitutes an excellent detector of
trap contamination, the observation that the finesse has not
degraded during the loading process confirms the cleanliness of
the loading technique.

\section{Optimizing the total number of ions inside the cavity
mode} \label{sec:optimization} 
\begin{figure}\sidecaption
\resizebox{1\hsize}{!}{\includegraphics*{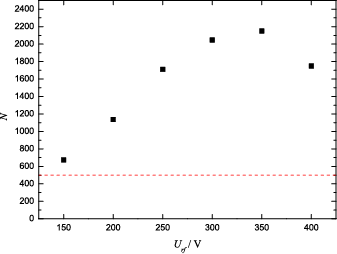}}
\caption{Number of ions, $N$, in the cavity mode volume versus
rf-voltage, $U_{rf}$. The interval from 150 V to 400 V corresponds
to ion densities between $6.8\times10^8-1.1\times10^9$ cm$^{-3}$.
The dashed line indicate the level above which the criterium for
strong collective coupling is potentially satisfied.} \label{Ncav}
\end{figure}

As mentioned in the introduction,
one of the main challenges for the realization of CQED with ion
Coulomb crystals is to load a large number of ions into the cavity
mode to enhance the collective atom-light coupling. For our trap,
with the optical cavity axis coinciding with the trap axis, the
highest number of ions in the cavity mode is achieved when the
product of the crystal length and the density, $\rho$, is
maximized. The length is controlled by both the end- and
rf-voltage, whereas the density is determined solely by the
rf-voltage (c.f. eq.~\ref{eq:density}). Ideally, we would
therefore wish to work at a very low $U_{end}$ and a very high
$U_{rf}$. In practice, we find that the crystals become unstable
at lengths above a few mm, although the exact length is very
dependent on the rf-voltage and on the total number of ions in the
crystal as well as the level of cooling power. In qualitative
terms, we interpret this as a result of rf-heating in the
crystal\cite{Blumel_89,Schiffer_00}. This heating drives the
radial motion of the ions, while the laser cooling beams act only
directly along the trap axis (see Fig.~\ref{setup}). Since the
coupling between axial and radial motion is weaker for longer than
for shorter prolate crystals, compensation of rf-heating by axial
laser cooling is less efficient in longer prolate crystals.
Increasing the cooling power may help to confine longer crystals
containing more ions, however, there is ultimately a tradeoff
between the maximal attainable length, the total number of
crystalized ions, and the ion density. Fig.~\ref{Ncav} shows the
maximal attainable number of ions in the cavity mode\footnote{$N$
is defined through the definition of the collective coupling
strength
$g_0\sqrt{N}=g_0\sqrt{\rho\int\Psi(\mathbf{r})^2d\mathbf{r}}$,
where the standing wave TEM$_{00}$ mode function is given by
$\Psi=\frac{w_0}{w(z)}e^{-(x^2+y^2)/w(z)^2}\mathrm{sin}^2(kz)$ and
where the integral is evaluated over the crystal length, $l$. With
this definition $N=\rho\frac{\pi w_0^2}{4}l$.}, $N$, as a function
of rf-voltage, $U_{rf}$, (density, $\rho$). Each point was found
by lowering the end-voltage and letting the crystal expand axially
until it became unstable and was lost. The number of ions in the
cavity mode is found to be maximal for rf-voltages around 350 V
($\rho\simeq8.3\times10^8$ cm$^{-3}$). Here, more than 2000 ions
are in the cavity mode, which would correspond to a collective
coupling strength of $g_0\sqrt{N}\simeq2\pi\times24$MHz. The
dashed line in the figure indicates the level above which the
number of ions becomes large enough to satisfy the strong
collective coupling criterium.

\section{Conclusion} \label{sec:conclusion}
We have achieved efficient \textit{in situ} loading of large ion
Coulomb crystals in a linear Paul trap with an integrated optical
resonator. This work demonstrates that such a high finesse optical
cavity can be incorporated into a linear Paul trap without
impeding the trapping of large ion Coulomb crystals, that these
crystals can have lengths comparable to that of the cavity (more
than one third of the cavity length) and that the finesse of the
cavity is preserved by
the method of resonant photoionization.\\
We have investigated the optimal trapping parameters for
maximizing the number of ions in the cavity mode and found that
more than 2000 ions could be confined in the cavity mode volume.
This number indicates that the system should be able to enter the
strong collective coupling regime and opens up for the
possibility of CQED experiments with ion Coulomb crystals.\\

%
\section*{Acknowledgements}
\label{sec:acknowledgements} The Authors would like to thank Joan
Marler and Magnus Albert for useful discussions and for
proofreading the manuscript.

This work has been financially supported by The Carlsberg
Foundation.
%

%
\bibliographystyle{unsrt}

\begin{thebibliography}{99}
\bibitem{Cirac_97}J. I. Cirac, P. Zoller, H. J. Kimble, H. Mabuchi,
Phys. Rev. Lett. \textbf{78}, 3221 (1997) 
\bibitem{Duan_01}L.-M. Duan, M. D. Lukin, J. I. Cirac, P. Zoller,
Nature \textbf{414}, 413 (2001) 
\bibitem{Berman} P. R. Berman (Ed.) \textit{Cavity Quantum
Electrodynamics}, Academic Press inc., London (1994)
\bibitem{Kimble_98} H. J. Kimble, Phys. Scr. \textbf{T76}, 127
(1998)
\bibitem{Boozer_06} A. D. Boozer, A. Boca, R. Miller, T. E. Northup, H. J.
Kimble, Phys. Rev. Lett. \textbf{97}, 083602 (2006)
\bibitem{Puppe_07} T. Puppe, I. Schuster, A. Grothe, A. Kubanek, K.
Murr, P.W.H. Pinkse, G. Rempe, Phys. Rev. Lett. \textbf{99},
013002 (2007)
\bibitem{Keller_2003} M. Keller, B. Lange, K. Hayasaka, W. Lange, H.
Walther, Appl. Phys. B \textbf{76}, 125 (2003) 
\bibitem{Mundt_2003} A.B. Mundt, A. Kreuter, C. Russo, C. Becher,
D. Leibfried, J. Eschner, F. Schmidt-Kaler, R. Blatt,
Appl. Phys. B \textbf{76}, 117 (2003) 
\bibitem{Mortensen_07}A. Mortensen, E. Nielsen, T. Matthey, M. Drewsen
J. Phys. B: At. Mol. Opt. Phys. \textbf{40}, F223 (2007) 
\bibitem{Mortensen_thesis} A. Mortensen, \textit{Aspects of Ion Coulomb Crystal based Quantum Memory for Light}, PhD
thesis, Department of Physics and Astronomy, University of Aarhus
(2005)
\bibitem{Herskind_proceedings} P. Herskind, A. Mortensen, J.L. S{\o}rensen, M. Drewsen,
\textit{Cavity-QED with ion Coulomb crystals}, in
\textit{Non-Neutral Plasma Physics Conference IV}, AIP Conference
Proceedings vol. 862, p. 292 (2006)
\bibitem{Coudreau_07} T. Coudreau, F. Grosshans, S. Guibal, L.
Guidoni, J. Phys. B: At. Mol. Opt. Phys. \textbf{40}, 413 (2007).
\bibitem{Prestage_89} J. D. Prestage, G. J. Dick, L. Maleki. J. Appl. Phys. \textbf{66}, 1013
(1989)
\bibitem{Drewsen_00} M. Drewsen, A. Br{\o}ner, Phys. Rev. A \textbf{62},
045401 (2000) 
\bibitem{Hornekaer_01} L. Hornek${\ae}$r, N. Kj${\ae}$rgaard, A. M. Thommesen, M. Drewsen,
Phys. Rev. Lett. \textbf{86}, 1994 (2001) 
\bibitem{Kjaergaard_00} N. Kj{\ae}rgaard, L. Hornek{\ae}r, A.M. Thommesen, Z. Videsen, M.
Drewsen, Appl. Phys. B \textbf{71}, 207 (2000) 
\bibitem{Brownnutt_07} M. Brownnutt, V. Letchumanan, G. Wilpers, R. C. Thompson, P. Gill, A. G.
Sinclair, Appl. Phys. B \textbf{87}, 411 (2007) 
\bibitem{Hendricks_07} R. J. Hendricks, D. M. Grant, P. F. Herskind, A. Dantan, M.
Drewsen, Appl. Phys. B \textbf{88}, 507 (2007) 
\bibitem{Gulde} S. Gulde, D. Rotter, P. Barton, F. Schmidt-Kaler, R.
Blatt, W. Hogervosrt, Appl. Phys. B \textbf{73}, 861 (2001) 
\bibitem{Deslauriers_06_1} L. Deslauriers, S. Olmschenk, D. Stick, W. K. Hensinger,
J. Sterk, C. Monroe, Phys. Rev. Lett. \textbf{97}, 103007 (2006) 
\bibitem{Devoe_02} R. G. DeVoe and C. Kurtsiefer, Phys. Rev. A \textbf{65},
063407 (2002) 
\bibitem{Mortensen_04} A. Mortensen, J. J. T. Lindballe, I. S. Jensen, P. Staanum, D. Voigt, M.
Drewsen, Phys. Rev. A \textbf{69}, 42502 (2004) 
\bibitem{Herskind_07} P. Herskind, J. Lindballe, C. Clausen, J. L. S{\o}rensen, M. Drewsen,
Opt. Lett. \textbf{32}, 268 (2007) 
\bibitem{Madsen_00} D. N. Madsen, S. Balslev, M. Drewsen, N.
Kj{\ae}rgaard, Z. Videsen, J. W. Thomsen, J. Phys. B: At. Mol. Opt \textbf{33}, 4981 (2000) 
\bibitem{Emsley_95} J. Emsley, \textit{The Elements} Oxford Chemistry Guides (Oxford Univ.
Press, New York, NY, 1995)
\bibitem{Nortershauser_98} W. N\"{o}rtersha\"{u}ser, K. Blaum, K. Icker, P. M\"{u}ller, A. Schmitt,
K. Wendt, B. Wiche, Eur. Phys. J. D 2, 33 (1998)
\bibitem{Blumel_89} R. Bl\"{u}mel, C. Kappler, W. Quint, H. Walther, Phys. Rev. A \textbf{40}, 808 (1989)
\bibitem{Schiffer_00} J. P. Schiffer, M. Drewsen, J. S. Hangst, L.
Hornek{\ae}r, Proc. Natl. Acad. Sci. \textbf{97}, 10697 (2000)
\end{thebibliography}

%
\end{document}